\documentclass[12pt,preprint]{aastex}

\shorttitle{On the flaring of jet-sustaining accretion disks} \shortauthors{F.
  Namouni}

\begin{document}        

\title{On the flaring of jet-sustaining accretion disks}

\author{Fathi Namouni}

\affil{CNRS, Observatoire de la C\^ote d'Azur, BP 4229, 06304 Nice, France}

\email{namouni@obs-nice.fr}         

\begin{abstract}
  Jet systems with two unequal components interact with their parent accretion
  disks through the asymmetric removal of linear momentum from the star-disk
  system. We show that as a result of this interaction, the disk's state of
  least energy is not made up of orbits that lie in a plane containing the
  star's equator as in a disk without a jet. The disk's profile has the shape
  of a sombrero curved in the direction of acceleration.  For this novel state
  of minimum energy, we derive the temperature profile of thin disks. The
  flaring geometry caused by the sombrero profile increases the disk
  temperature especially in its outer regions. The jet-induced acceleration
  disturbs the vertical equilibrium of the disk leading to mass loss in the
  form of a secondary wind emanating from the upper face of the disk.  Jet
  time variability causes the disk to radially expand or contract depending on
  whether the induced acceleration increases or decreases.  Jet time
  variability also excites vertical motion and eccentric distortions in the
  disk and affects the sombrero profile's curvature. These perturbations lead
  to the heating of the disk through its viscous stresses as it tries to
  settle into the varying state of minimum energy. The jet-disk interaction
  studied here will help estimate the duration of the jet episode in star-disk
  systems and may explain the origin of the recently observed one-sided
  molecular outflow of the HH 30 disk-jet system.
\end{abstract}


\section{Introduction}

Stellar jets interact dynamically with their parent accretion disk in two
fundamental ways. First, they extract angular momentum from the inner part of
the disk \citep{b8}, a theoretical prediction that has been confirmed recently
by the observation of velocity asymmetries in the jet of RW Auriga
\citep{b15,b7}.  Second, a stellar jet has been proposed as the physical
carrier of high temperature minerals from the Sun's vicinity to the outer
parts of the early solar system \citep{b16,b17,b18}. The discovery of
Calcium-Aluminum rich inclusions in the grains of Comet Wild 2 collected by
{\sc Stardust} seem to confirm this hypothesis. 

In this paper, we study a new dynamical interaction between the disk and jet
that arises from the asymmetric removal of linear momentum from the star-disk
system. A growing number of stellar jets appear to be asymmetric as the
ejection velocities of the jet and counterjet differ by about a factor of 2
\citep{b10,b11,b12,b14,b13}. Jet launching regions are confined to the inner
part of the disk with estimates from 0.01 AU for the X-wind model to a few AU
for disk-wind models (e.g. 1.6 AU for RW Auriga, \citealt{b15}). The
integrated momentum loss over the launching region accelerates the center of
mass of the star-disk system which coincides with the star's center for
axisymmetric systems. A gas element in the outer disk sees the matter around
which it revolves accelerating and responds dynamically to such change.  This
response is the object of this work.

An order of magnitude for the acceleration is obtained from the observed
mass loss rates and ejection velocities in disk-jet systems as:
\begin{equation}
A \sim  10^{-13}\, 
\left(\frac{\dot M}{10^{-8} M_\odot\, \mbox{\rm  yr}^{-1}}\right)\, 
\left(\frac{v_e}{300 \,\mbox{\rm km\,s}^{-1}}\right)\,
\left(\frac{M_\odot}{M}\right) \mbox{\rm km\,s}^{-2}. \label{acc-mag}
\end{equation}
where $M$ is the star's mass augmented by that of the jet launching region of
the disk, $\dot M$ is the mass loss rate, and $v_e$ is the jet's ejection
velocity \citep{b9}. This estimate is an instantaneous lower bound on the
total momentum loss as we lack long time span observations of stellar jets as
well as observations of the jet engine within a few AU from the star. Although
small, the acceleration amplitude (\ref{acc-mag}) was shown to offer a
possible explanation for the large eccentricities and secular resonances of
extrasolar planets' orbits provided the planets formed in a jet-sustaining
disk (\cite{b19} hereafter Paper I). Over the duration of the asymmetric jet
episode, the star-disk system acquires a residual velocity, $V$, that must be
smaller than the stellar velocity dispersion in the Galaxy,
$\left<v_g\right>$. This constraint yields an upper bound on the duration of
the asymmetric jet episode, $\tau$, as:
\begin{equation}
\tau \leq
10^5
 \ \frac{3\times
10^{-12}\,{\rm km\,s}^{-2}}{A}\ \frac{\left<v_g\right>}{10\, {\rm km\, s}^{-1
}} \, {\rm years}, \label{duration}
\end{equation}
where we used $V\sim A\tau$. Asymmetric momentum removal would contribute at
most a few km\,s$^{-1}$ to the observed velocity dispersion of young stars (of
order a few tens of km\,s$^{-1}$, \cite{veldisp}) leading to an upper bound on
the jet's lifetime comparable to the viscous time of the disk. The observation
of (the same) asymmetric jets since 1994, yields a lower bound of $\sim 10$
years on how long asymmetric jets may be sustained.

Jet asymmetry may have two origins: the ambient medium and/or an asymmetric
magnetic field that threads the disk. The former \citep{assym1} has the
advantage of not adding more complex geometries to the magnetic field
responsible for jet generation. The latter option of an asymmetric magnetic
field has been considered in the context of AGN systems where it has been
shown to warp the inner parts of the accretion disks through the asymmetric
local magnetic pressure \citep{referee}. Future observations inside a few AU
from the central star will help elucidate the origin of jet asymmetry.

In this work, we show that the acceleration of the star-inner-disk system with
respect to the outer disk (section 2) alters the disk's state of minimum
energy from a plane to a sombrero-shaped surface curved along the direction of
acceleration (section 3). This flaring geometry has new consequences for the
stellar irradiation of the disk (section 4) as well as for its vertical
structure (section 5). Jet time-variability is reflected in the varying
acceleration imparted locally to the disk leading to the global radial
expansion and contraction of the disk (section 6). These results are summed up
in section 7 by illustrating the disk's response as the jet is launched, and
discussing their possible application to the HH30 disk-jet system.

\section{Jet-induced acceleration}
\begin{figure}\begin{center}
\includegraphics[width=125mm]{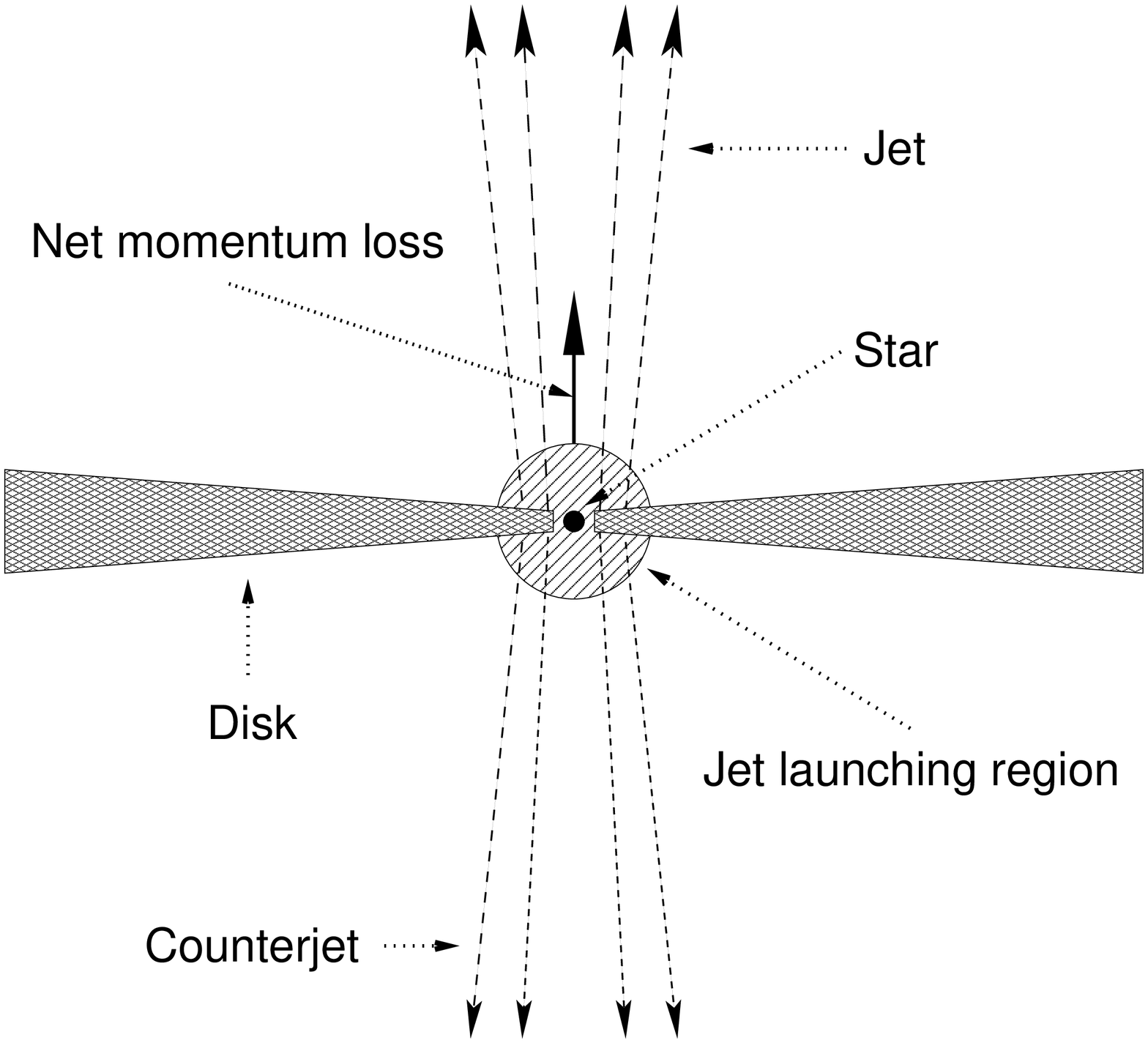}\end{center}
\caption{Schematic of the star-disk system. The momentum loss from the upper
  component is larger than that of the lower component. The net momentum loss
  from the jet launching region, which extends to a few astronomical units
  from the star, modifies the dynamics of the outer disk according to Eq.
  (\ref{motion}).}
\label{f1}
\end{figure}

Jet-induced acceleration arises from the removal of momentum from the
star-disk system (Fig.~\ref{f1}) in an inertial frame. To see this, we divide
the star-disk system into three particles: the star, an inner disk
materialized by a particle that looses mass and linear momentum located at 0.5
AU, and the outer disk materialized by a constant mass particle located far
outside, for instance at 50 AU. This is justified because the jet launching
region is confined to the inner part of the disk.  Gauss's theorem (the
divergence theorem) for the gravitational potential of the form $r^{-1}$
implies that the outer particle (outer disk) orbits around the combined masses
of the star and the inner particle (the inner disk) which form what we shall
call the central object.  The outer particle is not concerned by the details
of mass and momentum losses in the central object around which it revolves, it
only responds to them.  The central object or more accurately, the center of
mass of the star and the inner disk, is loosing mass and linear momentum. It
is therefore being accelerated with respect to the outer particle by an amount
equal to the ratio of the momentum loss rate from the inner particle divided
by the total mass of the central object --note that the inner particle itself
is accelerated by the ratio of the momentum loss rate to its mass. The
two-body problem, central object-outer particle, being symmetric, this also
amounts to saying that the outer particle is accelerated with respect to the
central object that remains stationary.  This shows that the outer particle
evolution may be described by the equation:
\begin{equation}
\frac{{\rm d} {\bf v}}{{\rm d} t}=-\frac{GM}{|{\bf x}|^3}\,{\bf
  x} +{\bf A}, \label{motion}
\end{equation}
where ${\bf x}$ and ${\bf v}$ are the position and velocity and $G$ is the
gravitational constant.  The mass $M$ is that of the star augmented by that of
the inner particle and corresponds to the total mass contained in the jet
launching region. The acceleration ${\bf A}$ is obtained as the ratio of the
total momentum loss to the total mass $M$.  In practice, we can neglect the
contributions of the disk's mass and its variations in $M$ for two reasons.
First, the mass of the inner disk is small compared to the mass of the star.
Second, the mass loss is small and amounts to a correspondingly small
dynamical effect given by the Jeans radial migration rate $\dot
r/r=10^{-8}$\,yr$^{-1}$ for $\dot M=10^{-8} M_\odot$\,yr$^{-1}$ \citep{jeans}.

From this simple picture, we can understand what goes on in an extended
continuous disk. Mass and linear momentum loss modify the disk's equations of
motion by including the disk's gravitational potential as well as the momentum
loss from the disk (in the case of disk winds). The contribution of the disk's
potential is negligible because the disk's mass is small and the migration
resulting from mass loss is also small. As for the momentum loss, it is a
function of the radius, $r$, that represents the integrated momentum loss up
to $r$. Calling $r_{\rm JLR}$ the outer radius of the jet launching region,
then for $r\geq r_{\rm JLR}$, the acceleration is not a function of $r$ as it
represents the total momentum loss from the star-disk system. For $r\leq
r_{\rm JLR}$, the variation with $r$ of the momentum loss depends on the model
of jet generation. Being interested in the flaring of the disk which mostly
concerns its outer parts, we may employ Eq.  (\ref{motion}) to describe the
evolution of a gas element located outside of the jet launching region where
$A$ is independent of $r$.

\section{State of minimum energy: The sombrero profile} 
We consider the steady state of the disk and accordingly take ${\bf A}$ to be
constant. The dynamical problem described by equation (\ref{motion}) has two
constants of motion, the energy $E=v^2/2-GM/|{\bf x}|-{\bf A}\cdot {\bf x}$,
and the projected component of the angular momentum ${\bf h}={\bf x}\times
{\bf v}$ on the direction of acceleration $h_z={\bf h}\cdot {\bf A}/A$.

\begin{figure}
\begin{center}
\includegraphics[width=127.5mm]{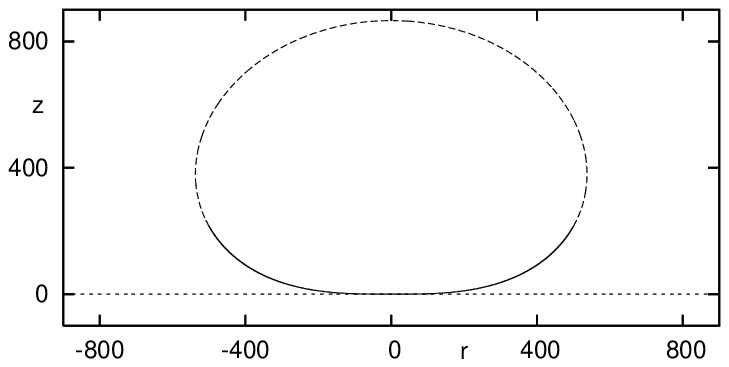}\\
\ \includegraphics[width=127.5mm]{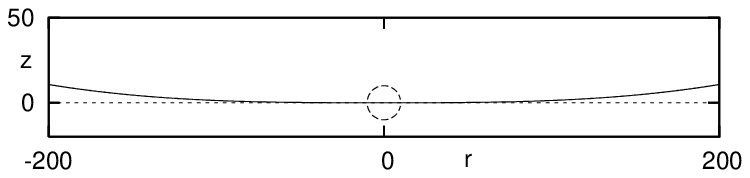}
\end{center}
\caption{
  The locus of circular orbits in the $rz$--plane (AU) for an acceleration
  $A=8\times 10^{-12}\ {\rm km}\, {\rm s}^{-2}$. The truncation radius $a_{\rm
    kplr}=500$\, AU. The upper panel shows all possible orbits: stable (solid
  line) and unstable (dashed line). The lower panel is a zoom of the stable
  orbits inside a radius of 200 AU around the star (dashed circle not to
  scale).}
\label{f2}
\end{figure}

Before discussing the situation where the jet is not perpendicular to the
disk, we first assume that the acceleration lies along the disk's plane
normal.  Least energy orbits make up the invariable surface of the disk -- if
there is no acceleration, this surface is the disk's midplane. We determine
the orbits of least energy by making use of cylindrical coordinates ($r$,
$\theta$, $z$) and by expressing the conservation of the vertical component of
angular momentum $h_z=r^2\dot \theta$. The energy equation then reads:

\begin{equation}
E=\frac{\dot r^2+\dot z^2}{2}+\frac{h_z^2}{2r^2}-\frac{GM}{\sqrt{r^2+z^2}}
-Az. \label{energy}
\end{equation}
Minimizing the energy integral yields $\dot r=\dot z=0$ which corresponds to
circular orbits and shows that $r$ and $z$ are not independent as the star's
pull balances the acceleration $A$ according to:
\begin{equation}
\frac{GM
  z}{(r^2+z^2)^{3/2}}=A.
\end{equation}
The vertical location of a circular orbit of radius $r$ and its orbital
rotation rate, $\Omega=\dot\theta$, are given as:
\begin{equation}
r=\left[\left(\frac{GM z}{A}\right)^\frac{2}{3}-z^2\right]^\frac{1}{2},
\ \ \ \Omega=
\left[\frac{GM}{\left(r^2+z^2\right)^{3/2}}\right]^\frac{1}{2}
\label{sombrero}. 
\end{equation}
We see that orbits of least energy are circular but do not lie in the
equator plane of the star, instead they hover above it in the direction
of acceleration.  The locus of circular orbits in the $rz$--plane is shown in
Fig.~(\ref{f2}, upper panel).  However, not all circular orbits are stable
in the long term. There is a natural truncation radius around the star where
the star's gravitational acceleration becomes comparable to $A$ (Paper I).
Beyond this radius, the star's pull is weak and orbits escape its gravity. The
location of this radius denoted $a_{\rm kplr}$ is given by the equality of the
acceleration excitation time $T_A=2\pi r\Omega/3 |A|$ (Paper I) and the
orbital period $T=2\pi/\Omega$. It is given as:
\begin{equation}
a_{\rm kplr}=\left[\frac{GM}{3|A|}\right]^\frac{1}{2}\simeq 10^3\,
\left[\frac{2\times 10^{-12}\ {\rm km}\, {\rm s}^{-2}}{|A|}\right]^\frac{1}{2}\, \left[\frac{M}{M_\odot}\right]^\frac{1}{2}\, {\rm AU}.\label{akplr}
\end{equation}
An accretion disk accelerated by a jet uses the stable orbits of the minimum
energy state. The profile thus obtained is that of a sombrero as shown in Fig.
(\ref{f2}, lower panel). We call $z=S(r)$ the stable disk profile obtained by
inverting Equation (\ref{sombrero}). We will also use the linearized profile,
$S(r)$ for $z\ll r$ which reads: $z=Ar^3/GM=r^3/3a_{\rm kplr}^2$.  We remark
that the sombrero profile coincides with the steady state profile of a thin
jet-sustaining disk. For (realistic) thick disks, the sombrero corresponds to
the invariable surface, the equivalent of the disk's midplane if there were no
acceleration, as well as to the disk's surface of maximum density (section 5).
In the case where the acceleration is not initially perpendicular to the disk,
the state of least energy is still a sombrero whose symmetry axis is that of
the acceleration. To reach this state, the disk experiences a transient phase
during which it dissipates energy (section 6).

\section{Temperature}
The stellar irradiation of a flat thin disk (i.e. a planar disk with zero
thickness) with no jet-induced acceleration yields a decreasing temperature
profile with distance \citep{b1}. It is given as $T_{\rm d}\propto T_*
(R_*/r)^{3/4}$ where $T_{\rm d}$ is the disk's temperature, $T_*$ and $R_*$
are the star's temperature and radius respectively.  For a non-planar disk,
the disk's temperature at distance $r$ from the star is related to the stellar
temperature through:
\begin{equation}
T^4_{\rm d}=T_*^4
   \ \left[\frac{rz^\prime-z}{2r}+\frac{2R_*}{3\pi r}
   \right]\ \left[\frac{R_*}{r}\right]^2, \label{eqn11}
\end{equation}
\citep{b3} where $z$ is the vertical location of the disk and
$z^\prime={\rm d}z/{\rm d}r$. For $z\ll r$, we substitute the linearized
sombrero profile $z=S(r)$ into (\ref{eqn11}) to find:
\begin{equation} 
T_{\rm d}=  T_* \left[\frac{R_*^2}{3 a_{\rm kplr}^2} +\frac{2R_*^3}{3\pi
    r^3}\right]^\frac{1}{4}.
\end{equation}
The outer regions ($r\gg R_*$) are those that contribute most to the
temperature profile. Accordingly, the disk's temperature is given as:
\begin{equation} 
T_{\rm d}\simeq  13\,  \left[\frac{T_*}{4000\,\mbox{K}}\right]\
                       \left[\frac{R_*}{2 R_\odot}\right]^\frac{1}{2}\
                       \left[\frac{500\,\mbox{AU}}{a_{\rm kplr}}\right]^\frac{1}{2} \ \mbox{K}.\label{td}
\end{equation}
The fundamental state of the disk therefore has a residual temperature
regardless of radius. The sombrero profile cancels out the $r^{-2}$ distance
factor responsible for the decrease of temperature with distance. Relaxing
$z\ll r$ and using the full shape of $S(r)$ leads to a modest temperature
increase with distance.  The increase and decrease of the slopes of $z/r$ on
both sides of the sombrero for a finite thickness disk will make the
temperature grow beyond the estimate of Eq.  (\ref{td}) for the upper side of
the disk and exposes the lower side to some irradiation.  The unequal heating
of jet-sustaining disks therefore sets a local vertical temperature gradient
that leads to the stratification of the disk.

\section{Hydrostatic equilibrium and wind generation}
For a disk with a finite thickness, we may seek the vertical profile obtained
from hydrostatic equilibrium.  The hydrostatic equilibrium equations are now
modified by the presence of the vertical acceleration ${\bf A}$.  Assuming an
isothermal disk of pressure $p=c^2\rho$ where $\rho$ is the density and $c$
the sound speed, the hydrostatic equilibrium equation reads:
\begin{equation}
\frac{{\rm d}p}{{\rm d}z} +\frac{G Mz\rho}{\left(r^2+z^2\right)^{3/2}}-\rho
A=0, \label{hydroeq}
\end{equation}
whose solution is :
\begin{equation}
\rho=\rho_0(r)\exp\frac{1}{c^2} \left[\frac{G M}{\sqrt{r^2+z^2}}+Az\right].\label{hydro}
\end{equation}
The sombrero profile here corresponds to the surface of maximum density. With
respect to this profile and where $z\ll r$, the disk's thickness,
$H=c/\Omega$, is similar to that of an unaccelerated disk. This solution
however has a flaw: the linear term related to the acceleration yields an
infinite disk mass when $\rho$ is integrated over all space.  The finite value
of $\rho$ for large positive $z$ is indicative of a matter outflow from the
upper side of the disk in response to the lowering of the vertical pull of the
star by the acceleration ${\bf A}$. Consequently, Eq.  (\ref{hydroeq}) can no
longer describe the vertical state of the disk.  The steady state properties
of the new wind component may be described by the unmagnetized Grad-Shafranov
equation \citep{b21} subject to the finite duration acceleration ${\bf A}$. We
expect the density and velocity profile of the wind component to bear some
resemblance to the hydrodynamic jets generated as force-free vortex funnels in
accretion disks \citep{b21x}.

\section{Variability}
\subsection{Vertical evolution}
Stellar jets are time-variable processes \citep{var1, var4,var2, var3, b7}. To
ascertain the consequences of the jet-induced acceleration's time variations,
we first consider an initially circular planar disk subject to a vertical
acceleration of magnitude $A$. This models an instantaneous increase of
acceleration from zero to $A$.  Solving the vertical equation of motion in
(\ref{motion}) under the assumption that $z\ll r$ yields:
\begin{equation}
z=\frac{Ar^3}{GM}\ (1-\cos\Omega t). \label{vertical}
\end{equation}

\begin{figure}\begin{center}
\includegraphics[width=127.5mm]{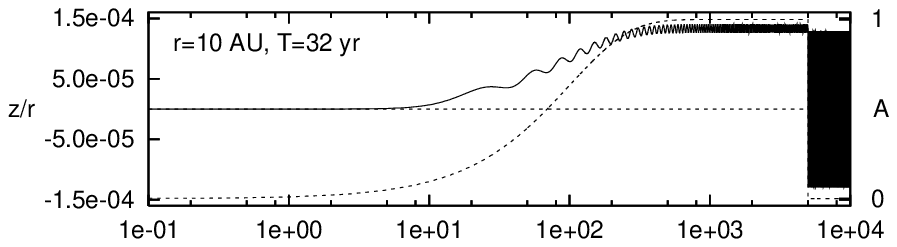}\\
\includegraphics[width=127.5mm]{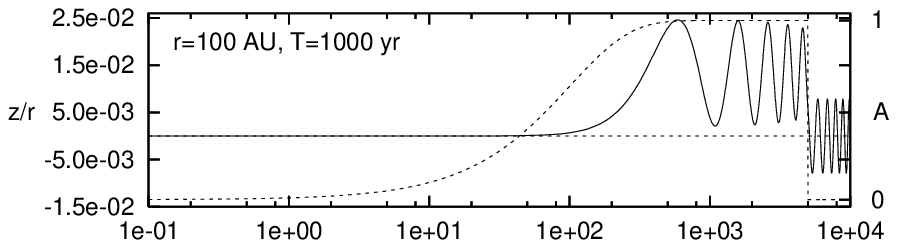}\\
\includegraphics[width=127.5mm]{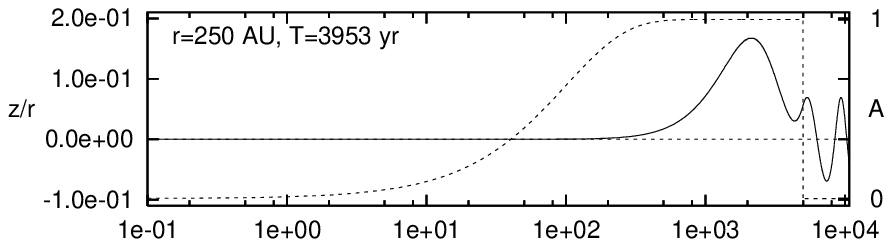}\\
\includegraphics[width=127.5mm]{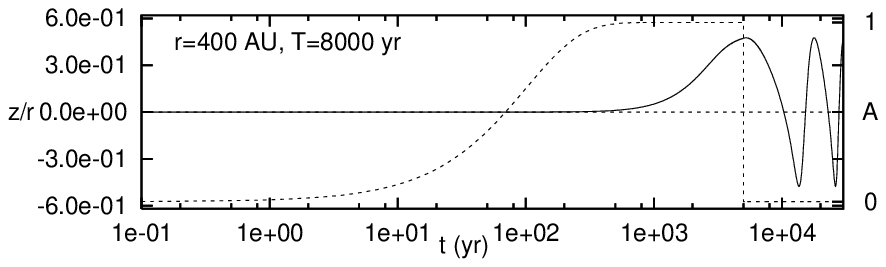}\end{center}
\caption{Vertical evolution of the outer disk. The equations of motion
  (\ref{motion}) were integrated numerically to simulate the vertical motion
  of a gas element initially located in the disk's midplane before the
  jet-induced acceleration is activated. The acceleration is modeled as
  $A=8\times 10^{-12}\ {\rm km}\, {\rm s}^{-2}\times (1-\exp[-t/100\,{\rm
    yr}])\times W(t)$ where $W(t)=1$ if $0\leq t\leq 5000$\,yr and zero
  otherwise. The four panels (left scale) illustrate $z(t)/r$ for $r=$10\,AU,
  100\,AU, 250\,AU and 400\,AU --the corresponding dynamical times $T$ are
  given in each panel. The dashed-line profile (right scale) in the four
  panels shows the acceleration $A$ normalized to its maximum value $8\times
  10^{-12}\ {\rm km}\, {\rm s}^{-2}$.}
\label{f3}
\end{figure}

The vertical motion is an oscillation with respect to the linearized sombrero
profile $S(r)$ on a timescale equal to the local orbital period in the disk,
$T$.  The oscillation amplitude is given by the distance from the initial
state which in this case is the equator plane to the location of the sombrero
profile.  If the acceleration grew from zero to $A$ on a timescale, $|A/\dot
A|$, longer than the orbital period, $T$, the disk's midsurface height $z$ would
follow the evolving sombrero profile adiabatically and there will be no
oscillation with respect to the surface of least energy. To illustrate the two
excitation regimes, sudden and adiabatic, we show in Fig.~(\ref{f3}) the
response of the disk's midsurface height obtained by numerically integrating the
equation of motion (\ref{motion}). The acceleration's time dependence is taken
as $A=A_{\rm max}(1-\exp[-t/100\,{\rm yr}])\times W(t)$ where $A_{\rm
  max}=8\times 10^{-12}\ {\rm km}\, {\rm s}^{-2}$, the same magnitude as in
Fig.~(\ref{f2}), and $W(t)$ is a window function that vanishes outside the
range [0,5000\,yr] and equals unity inside it. The minimum truncation radius
is therefore 500\,AU. As the launching timescale is $|A/\dot A|=100$\,yr, the
disk evolves adiabatically to the sombrero profile inside 22\,AU as can be
seen on $z(t)/r$ at $r=10$\,AU ($T=32$\,yr) in Fig.~(\ref{f3}). Outside
22\,AU, the dynamical time $T$ is longer than the launching timescale of
100\,yr implying that jet launching is perceived as a sudden excitation by
most of the outer disk whose typical response is described by Eq.
(\ref{vertical}) and shown in Fig (\ref{f3}) for $r=100$\,AU.

The acceleration is shut off suddenly after 5000\,yr. This determines an outer
limit, $r\sim 300$\,AU, in the disk beyond which the state of least energy
remains the disk's initial state (star's equator plane) as the dynamical time
outside this radius is too long to travel all the way to the location of the
sombrero profile (Eq.  \ref{vertical}).  Beyond this radius, the jet-induced
acceleration episode is felt as a short (with respect to the orbital period)
velocity pulse to the outer disk resulting in the excitation of vertical
motion with respect to the disk's initial midplane as seen in Fig.~(\ref{f3})
for $r=400$\,AU. This type of excitation has been discussed extensively in
Paper I (section 6). The sudden absence of acceleration is felt everywhere in
the disk, where for instance at $r=10$\,AU, the lack of the jet-induced
acceleration that balanced the star's vertical gravitational pull leads to a
vertical oscillation with respect to the equator plane and with an amplitude
equal to the distance from the sombrero to that plane. This effect is the same
as the excitation described by Eq.  (\ref{vertical}) only this time the
invariable surface is the disk's midplane and the initial state is the sombrero
profile.  For both the accelerated and unaccelerated phases, the vertical
oscillations with respect to the surface of least energy (be it a sombrero if
$A\neq 0$ or a plane if $A=0$) are damped by the viscous stresses of the disk.
For instance, in a more realistic simulation taking into account the disk's
viscosity, the observed oscillations about the sombrero height $z/r\sim
10^{-2}$ for $r=100$\,AU, would be erased and the disk would settle into the
sombrero profile before the acceleration is shut off at $t=5000$\,yr.  We note
that the illustration of the disk's midsurface height response to the
acceleration's growth shown here is not peculiar to the launching of the jet
and equally represents the disk's response to more general variations of the
jet-induced acceleration as those associated with a varying ejection velocity
at the source \citep{var1,var2}.

\subsection{Horizontal evolution}
For a time-variable acceleration, the conservation of the specific angular
momentum $h_z=r^2\Omega$ forces the mean orbital radius to change in order to
compensate for the increase of the midsurface height $z$ to the location of
sombrero profile.  The rate of this change is found using the conservation of
$h_z$ only as:

\begin{equation}
\dot r=3 \ \frac{z}{r}\ \dot z, \label{hz}
\end{equation} 
where we neglected a term of second order in $z/r$. The rate~(\ref{hz}) is
valid regardless of the time dependence of $A$ and whether the initial state
of the disk is planar. The change in radius is of second order in $A$ and
leads to a radial expansion of the disk if the acceleration increases with
time. If however $A$ decreases, the midsurface height $z$ is lowered to the
new sombrero location and Eq.  (\ref{hz}) shows that the disk undergoes a
radial contraction. The expansion and contraction occur on the local dynamical
time $T$ if it is larger than the variability timescale $|A/\dot A|$ (as in
Eq.  \ref{vertical}). Otherwise, radial motion occurs adiabatically with
respect to variability. Outside the radius where $T=|A/\dot A|$, it is the
former evolution that takes place implying that contraction and expansion
start from the inside of the disk and proceed outward (as $T$ increases with
$r$). The local radial displacement $\Delta r$ is given as:
\begin{equation} 
\frac{|\Delta r|}{r}= \frac{3\epsilon A^2r^4}{G^2M^2}=\frac{\epsilon}{3}\,
 \left[\frac{r}{a_{\rm kplr}}\right]^4 \label{exp-con}
\end{equation}
where $\epsilon=1$ if the acceleration's change is sudden, $|A/ \dot A|\leq
T$. For an adiabatic change, $|A/\dot A|\gg T$, the averaging of the
linearized equations of motion (\ref{motion}) yields $\epsilon=2/3$.  The
outer half of the disk is the most affected by the expansion and contraction:
at the disk's median radius, $r=a_{\rm kplr}/2$, $|\Delta r|/r\sim 2\%$. We
also note that this rate is much larger than that of the Jeans migration
(section 2).

\begin{figure}\begin{center}
\includegraphics[width=127.5mm]{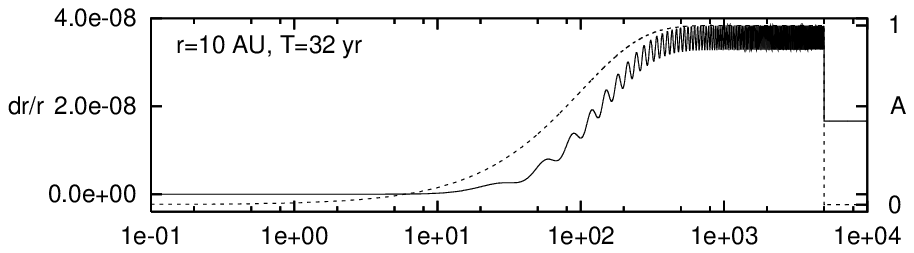}\\
\includegraphics[width=127.5mm]{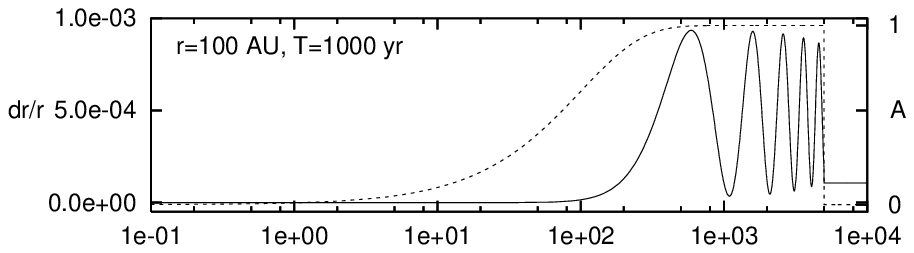}\\
\includegraphics[width=127.5mm]{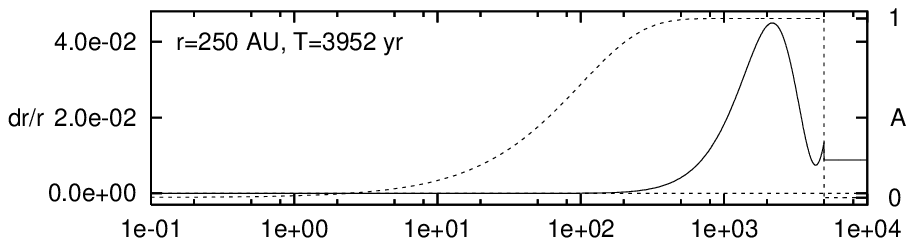}\\
\includegraphics[width=127.5mm]{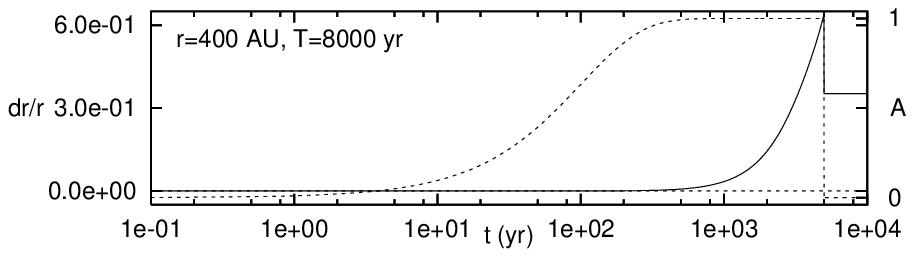}\end{center}
\caption{Radial expansion and contraction of the outer disk. 
  The equations of motion (\ref{motion}) were integrated numerically to
  simulate the radial motion of a gas element initially located in the disk's
  midplane before the jet-induced acceleration is activated. The acceleration
  is modeled as $A=8\times 10^{-12}\ {\rm km}\, {\rm s}^{-2}\times
  (1-\exp[-t/100\,{\rm yr}])\times W(t)$ where $W(t)=1$ if $0\leq t\leq
  5000$\,yr and zero otherwise. The four panels (left scale) illustrate
  $\Delta r/r$ for $r=$10\,AU, 100\,AU, 250\,AU and 400\,AU --the
  corresponding dynamical times $T$ are given in each panel. The dashed-line
  profile (right scale) in the four panels shows the acceleration $A$
  normalized to its maximum value $8\times 10^{-12}\ {\rm km}\, {\rm
    s}^{-2}$.}
\label{f4}
\end{figure}

Figure~(\ref{f4}) shows the local expansion and contraction of the disk at
different radii under the same conditions as those of the previous section and
Fig.~(\ref{f3}).  At $r=10$\,AU the expansion is adiabatic as the launching
timescale $|A/\dot A|$ is larger than the dynamical time, while the
contraction is sudden as the acceleration is abruptly turned off. Where the
dynamical timescale is much larger than the acceleration's duration as for
$r=400$\,AU, there is a net radial expansion outward as the whole acceleration
episode is felt as a velocity kick to the outer disk (Paper I, section 6) ---
note for instance how for $r=400$\, AU, the expansion appears to be delayed
until the end of the acceleration phase.
 
Another consequence of the acceleration's variability is the local excitation
of eccentricity in the disk. Using the linearized equations of motion
(\ref{motion}) for the radius $r$, it can be shown that the local eccentricity
growth rate due to the vertical acceleration is:
\begin{equation}
\dot e= \frac{3A^2r^4\Omega}{\pi G^2M^2},\label{ecc-rate}
\end{equation}
with a secular turnaround time equal to $T_A$. The maximum eccentricity is
therefore $ e_{\rm max}= (r/a_{\rm kplr})^2/3$.  We note that the eccentricity
gradient ${\rm d} (re)/{\rm d}r=(r/a_{\rm kplr})^2$ is always smaller than
unity implying that the forced eccentric distortions of the gas streamlines do
not make them cross and consequently they do not shock.  The estimate of Eq.
(\ref{ecc-rate}) is valid during the acceleration phase. Outside the location
where the dynamical time is smaller than the duration of the acceleration
phase, the net effect of acceleration amounts to a small amplitude velocity
kick imparted to the disk. To the fluid elements of the disk's midplane, the
velocity kick imparts an eccentricity given as (Paper I, section 6):
\begin{equation}
e(r)= \frac{V^2r/GM}{1+V^2r/GM},\label{ecc-rateout}
\end{equation}
where $V$ is the total velocity obtained by integrating the acceleration over
its duration.  The global response of the disk's viscous stresses to the
eccentric disturbance (\ref{ecc-rate}) depends on the prescription of the
viscous stresses that is employed as well as on the density profile
\citep{b51}.  We expect the viscous stresses to damp the eccentric distortions
as well as the vertical oscillations leading to the heating of the disk. In
the outermost part of the accretion disk where the dynamical time is long
compared to the acceleration's duration and where the viscous timescale is
similarly long, we expect jet-sustaining accretion disks to exhibit global
eccentric distortions.

\section{Synthesis}
Stellar jets are ubiquitous processes of star formation and star-disk
interaction.  The asymmetric removal of linear momentum from the star-disk
system accelerates the outer part of the disk with respect the jet launching
region. The acceleration amplitude is small (\ref{acc-mag}) and can be modeled
as a time dependent vector directed along the star's rotation axis. The state
of minimum energy of the disk is a sombrero-shaped surface made up of circular
planar orbits that hover above the star's equator plane (\ref{sombrero})
increasingly higher with distance. As the jet is launched, the initially
circular equator disk moves towards the state of minimum energy and the
truncation radius moves from infinity to the location given by (\ref{akplr}).
If the disk's size is larger than the minimum truncation radius over the
acceleration phase, the disk will be truncated by the jet-induced
acceleration.  There is a specific radius in the disk where the orbital period
$T$ matches the jet launching timescale $|A/\dot A|$.  Inside this radius, the
disk moves away from the equator plane to the instantaneous sombrero surface
adiabatically. Outside the matching radius, vertical motion (\ref{vertical})
as well as eccentricity distortions (\ref{ecc-rate}) with respect to the
sombrero are excited. The conservation of angular momentum implies that as the
jet is launched the disk expands radially (\ref{hz},\ref{exp-con}).  Viscous
stresses damp the vertical motion and the eccentricity distortions to heat up
the disk especially in its outer regions.  The steady state of the disk thus
assumes a sombrero profile that exposes the upper face of the disk to the
star's light canceling at first order the irradiation's decrease with distance
(\ref{td}). The disk cannot maintain hydrostatic equilibrium under the jet
acceleration (\ref{hydro}).  Instead the disk looses mass from its upper side
giving rise to a secondary asymmetric wind component. When the jet-induced
acceleration decreases, the sombrero looses curvature, the disk contracts
radially at the rate given by (\ref{exp-con}) and the truncation radius moves
outward (\ref{akplr}).

The mechanisms discussed above only require the onset of asymmetric momentum
removal in the star-disk system. Whether momentum is removed from the star
itself through some peculiar stellar wind component, or more realistically
from the disk or the star-disk interface \citep{assym1}, the outer disk will
respond to the integrated momentum loss from the matter around which it
gravitates.  The mechanisms discussed above also do not have a time
requirement on the acceleration's duration. The evolution towards the
sombrero, the radial contraction and expansion proceed from the inner part of
the disk outwards. Therefore, these mechanisms affect the disk inside the
radius where the dynamical time is comparable to the acceleration's duration.
In this sense, disk observations can be used to constrain the duration of the
jet episode and/or asymmetric momentum removal by identifying the outermost
radius where relaxation toward the sombrero profile has taken place.
Asymmetric momentum removal imprints two more signatures on the star-disk
system: the presence of a global eccentric distortion in the outermost part of
the disk, and an enhancement of the random component of the stellar velocity
in the Galaxy. The eccentric distortion results from a perceived sudden pull
of the star-inner disk system with respect to the outer disk along the jet
system's axis and should be an increasing function of radius
(\ref{ecc-rateout}) that does not depend on the acceleration's timescales.
The change in the stellar velocity dispersion may be ascertained by
statistically analyzing young stars' proper motion at different epochs of
their evolution.

A good illustration of some of the new properties discussed above is found in
the HH 30 disk-jet system. This system \citep{hh1} is a nearly edge-on $420$
AU disk that appears as two reflection nebulosities separated by a dark lane.
The jet and counterjet appear to be asymmetric although both the ejection
speed and the mass loss rate remain uncertain \citep{hh1,hh6}. The jet
exhibits variability in the form of knots with periods of 2.5 years
\citep{hh1}. The disk's variable lateral asymmetries appear mainly on one side
of the disk with possible periods from a few days to a few years \citep{hh2,
  hh7, hh5}.  More recently, \cite{hh3} reported the observation of a
molecular outflow emanating from the upper nebulosity. The outflow was traced
back to a scale smaller than the size of the disk while no rotation signatures
were detected above 1 km\,s$^{-1}$ at 200\, AU. The variable lateral
asymmetries are currently explained by the presence of a single hot spot on
the star requiring a complex magnetic field structure that differentiates
substantially between the two stellar poles \citep{hh2}. If the one-sided
outflow resulted from the entrainment of surrounding material then the cloud
structure would be substantially different on both sides of the disk.  In the
context of the jet-disk interaction studied here, it is natural to attribute
the lack of lateral asymmetries in the lower nebulosity to a change in the
disk's mean profile, and to identify the one-sided molecular outflow as the
new secondary wind component.  Further modeling of the disk temperature, wind
density and velocity profiles produced by the jet-induced acceleration will
help determine the properties of the HH 30 disk and its molecular outflow.

\acknowledgments The author thanks the referee for constructive comments that
helped improve the clarity of the paper and Christiane Froeschl\'e for
stimulating discussions and the careful reading of the manuscript.  This work
was supported by the Programme National de Plan\'etologie.

\clearpage

\clearpage

\clearpage

\clearpage

\end{document}